\documentstyle[aps,prb,twocolumn,epsf,floats]{revtex}
\begin{document}
\draft
\wideabs{
\title{Two-Dimensional Electron-Hole Systems in a Strong Magnetic Field:
       \protect\\
       Composite Fermion Picture for Multi-Component Plasmas}
\author{
   Arkadiusz W\'ojs$^{1,2}$, 
   Izabela Szlufarska$^{1,2}$, 
   Kyung-Soo Yi$^{1,3}$, and
   John J. Quinn$^1$}
\address{
   $^1$Department of Theoretical Physics, 
       University of Tennessee, Knoxville, Tennessee 37996, USA \\
   $^2$Institute of Physics, 
       Wroclaw University of Technology, Wroclaw 50-370, Poland \\
   $^3$Physics Department, 
       Pusan National University, Pusan 609-735, Korea}

\maketitle

\begin{abstract}
   Electron-hole systems on a Haldane sphere are studied by exact 
   numerical diagonalization.
   Low lying states contain one or more types of bound charged 
   excitonic complexes $X_k^-$, interacting through appropriate 
   pseudopotentials.
   Incompressible ground states of such multi-component plasmas are 
   found.
   A generalized multi-component Laughlin wavefunction and composite 
   Fermion picture are shown to predict the low lying states of an 
   electron-hole gas at any value of the magnetic field.
\end{abstract}

}

\paragraph*{Introduction.}
Recently there has been considerable interest in two dimensional systems
containing both electrons and holes in the presence of a strong magnetic 
field.\cite{kheng,shields,chen,wojs1,rashba,palacios,wojs2,lerner}
In such systems, neutral ($X^0$) and charged excitons ($X^-$) and larger 
exciton complexes ($X_k^-$, $k$ neutral $X^0$'s bound to an electron) 
can occur.
The excitonic ions $X_k^-$ are long-lived Fermions,\cite{palacios,wojs2} 
whose energy spectra contain Landau level structure.\cite{wojs1,wojs2}
In this paper we investigate by exact numerical diagonalization small
systems containing $N_e$ electrons and $N_h$ holes ($N_e\ge N_h$), 
confined to the surface of a Haldane sphere.\cite{haldane1}
For $N_h=1$ these systems serve as simple guides to understanding
photoluminescence.\cite{kheng,shields,chen,wojs1,rashba}
For larger values of $N_h$ it is possible to form a multi-component plasma
containing electrons and $X_k^-$ complexes.\cite{wojs2}
We propose a model \cite{halperin} for determining the incompressible 
quantum fluid states\cite{laughlin} of such plasmas, and confirm the 
validity of the model by numerical calculations.
In addition, we introduce a new generalized composite Fermion (CF) 
picture\cite{jain} for the multi-component plasma and use it to predict 
the low lying bands of angular momentum multiplets for any value of the 
magnetic field.

\paragraph*{Bound States.}
In a sufficiently strong magnetic field, the only bound electron-hole 
complexes are the neutral exciton $X^0$ and the spin-polarized charged 
excitonic ions $X_k^-$ (electron $e^-\equiv X_0^-$, charged exciton 
$X^-\equiv X_1^-$, charged biexciton $X_2^-$, etc.).
\cite{palacios,wojs2}
All other complexes found at weaker magnetic fields (e.g. spin-singlet 
charged exciton\cite{kheng} or spin-singlet biexciton) unbind.\cite{lerner}
The angular momenta of complexes $X^0$ and $X_k^-$ on a Haldane 
sphere\cite{haldane1} with monopole strength $2S$ are $l_{X^0}=0$ and 
$l_{X_k^-}=|S|-k$.\cite{wojs2}
The binding energies of an exciton, $\varepsilon_0=-E_{X^0}$, and of 
excitonic ions, $\varepsilon_k=E_{X_{k-1}^-}+E_{X^0}-E_{X_k^-}$ ($E_A$ 
is the energy of complex $A$) are listed in Tab.~\ref{tab1} for several 
different values of $2S$.
\begin{table}
\caption{
   Binding energies $\varepsilon_0$, $\varepsilon_1$, $\varepsilon_2$, 
   and $\varepsilon_3$ of $X^0$, $X^-$, $X_2^-$, and $X_3^-$, 
   respectively, in the units of $e^2/\lambda$.}
\begin{tabular}{rcccc}
 $2S$ & $\varepsilon_0$ & $\varepsilon_1$ 
      & $\varepsilon_2$ & $\varepsilon_3$ \\ \hline
  10 & 1.3295043 & 0.0728357 & 0.0411069 & 0.0252268 \\ 
  15 & 1.3045679 & 0.0677108 & 0.0395282 & 0.0262927 \\ 
  20 & 1.2919313 & 0.0647886 & 0.0381324 & 0.0260328
\end{tabular}
\label{tab1}
\end{table}
It is apparent that $\varepsilon_0>\varepsilon_1>\varepsilon_2>
\varepsilon_3$.
Depending on the ratio $N_e\!:\!N_h$, we expect to find different 
combinations of complexes that have the largest total binding energy.
When $N_e=N_h$ we expect $N_h$ neutral excitons $X^0$ to form.
When $N_e\ge2N_h$ the low lying states will contain $N_h$ charged 
excitons $X^-$ and $N_e-2N_h$ free electrons $e^-$.
For $N_h<N_e<2N_h$ we expect to find larger charged exciton complexes.

\begin{figure}[t]
\epsfxsize=3.35in
\epsffile{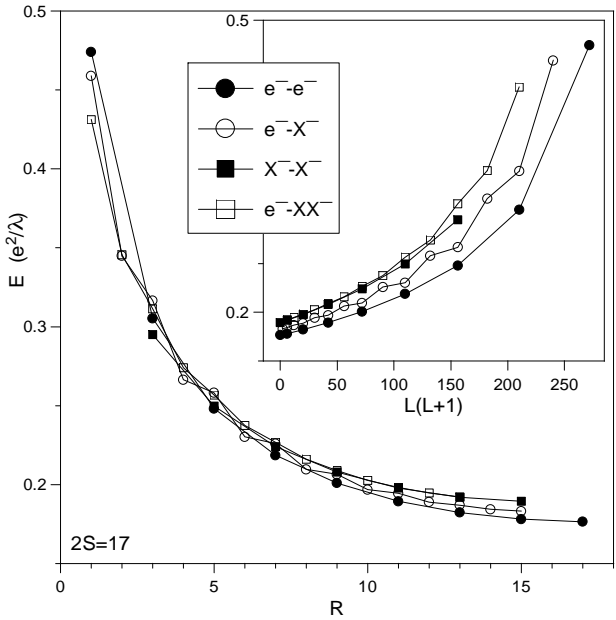}
\caption{
   Pseudopotentials $V_{e^-e^-}$ (filled circles), $V_{e^-X^-}$ (open
   circles), $V_{X^-X^-}$ (filled squares), and $V_{e^-X_2^-}$ (open 
   squares) on a Haldane sphere with $2S=17$, as a function of ${\cal 
   R}$ (main frame) and $L(L+1)$ (inset).
}
\label{fig1}
\end{figure}

\paragraph*{Pseudopotentials.}
Whether the states with largest binding energy form the lowest energy 
band of the electron-hole system depends on the interaction between
charged complexes $X_k^-$.
The interaction of a pair of charged particles $A$ and $B$ of angular 
momentum $l_A$ and $l_B$ can be described by a pseudopotential 
$V_{AB}(L)$ where $\hat{L}=\hat{l}_A+\hat{l}_B$ is the total pair angular 
momentum.\cite{haldane2}
It is convenient to plot pseudopotentials as a function of the relative 
angular momentum ${\cal R}=l_A+l_B-L$.\cite{wojs3}
Fig.~\ref{fig1} shows $V_{AB}({\cal R})$ for the pairs $e^-e^-$, $e^-X^-$, 
$X^-X^-$, and $e^-X_2^-$, at the monopole strength $2S=17$.
Roughly, the pseudopotential parameters $V_{AB}({\cal R})$ calculated 
for different pairs $AB$ and for a given $2S$ lie on the same curve.
Small differences between energies $V_{AB}$ calculated for different 
pairs at the same ${\cal R}$ are due to different values of $l_A$ and 
$l_B$ and to the finite size and polarization of composite particles.
Only the latter effect, important at small ${\cal R}$, persists for
$2S\rightarrow\infty$, i.e. in the planar geometry.

The major and critical difference between four plotted pseudopotentials 
lies in the allowed values of ${\cal R}$.
If all $A$ and $B$ were point charges, the allowed pair angular momenta 
for two identical Fermions ($A=B$) would be $L=2l_A-j$, where $j$ is an 
odd integer, i.e. ${\cal R}=1$, 3, \dots\ and ${\cal R}\le2l_A$.
For two distinguishable particles ($A\ne B$), the values of $L$ would 
satisfy $|l_A-l_B|\le L\le l_A+l_B$, i.e. ${\cal R}=0$, 1, 2, \dots and
${\cal R}\le2\min(l_A,l_B)$.
However, if $A$ or $B$ is a composite particle, one or more pair states 
with largest $L$ (smallest ${\cal R}$) are forbidden, and the corresponding 
pseudopotential parameters are effectively infinite ($AB$ repulsion has 
a hard core).
For $A=X_{k_A}^-$ and $B=X_{k_B}^-$, the smallest allowed ${\cal R}$ 
can be deduced from the mapping\cite{lerner} between the electron-hole 
and two-spin systems,
\begin{equation}
   {\cal R}_{AB}^{\rm min}=2\min(k_A,k_B)+1.
\label{eq1}
\end{equation}
Thus, in Fig.~\ref{fig1}, ${\cal R}_{e^-X^-}\ge1$, ${\cal R}_{X^-X^-}\ge3$, 
etc.

Low lying states of a system of $N_e$ electrons and $N_h$ holes can 
contain a number of charged complexes $X_k^-$ ($X^-$ and possibly larger 
ones) interacting with one another and with electrons through appropriate 
pseudopotentials.
It has been shown\cite{wojs3} that the Laughlin $\nu=1/m$ state occurs 
in the gas of (identical) Fermions if the pseudopotential increases faster 
than linearly as a function of $L(L+1)$ in the vicinity of ${\cal R}=m$.
As seen in the inset in Fig.~\ref{fig1}, this is true for both $V_{e^-e^-}$ 
and $V_{X^-X^-}$, and also (at even values of ${\cal R}$) for $V_{e^-X^-}$ 
and $V_{e^-X_2^-}$.
In Ref.~\onlinecite{wojs2} we found Laughlin states of one-component $X^-$ 
gas formed at $N_e=2N_h$.
In the present note we concentrate on a more general situation, where 
more than one kind of charged particles occur in an electron-hole system,
and find incompressible fluid states of such multi-component plasma.

\paragraph*{Numerical Results.}
As an illustration, we present first the results of exact diagonalization 
performed for the system with $N_e=8$ and $N_h=2$.
We expect low lying bands of states containing the following combinations
of complexes: (i) $4e^-+2X^-$, (ii) $5e^-+X_2^-$, (iii) $5e^-+X^-+X^0$, 
and (iv) $6e^-+2X^0$.
All groupings (i)--(iv) contain an equal number $N=N_e-N_h$ of singly 
charged complexes, however, both the angular momenta of involved complexes 
and the relevant hard cores are different.
The total binding energies are:
$\varepsilon_{\rm i}=2\varepsilon_0+2\varepsilon_1$, 
$\varepsilon_{\rm ii}=2\varepsilon_0+\varepsilon_1+\varepsilon_2$,
$\varepsilon_{\rm iii}=2\varepsilon_0+\varepsilon_1$, and 
$\varepsilon_{\rm iv}=2\varepsilon_0$.
Clearly, $\varepsilon_{\rm i}>\varepsilon_{\rm ii}>\varepsilon_{\rm iii}>
\varepsilon_{\rm iv}$.
However, which of the groupings contains the (possibly incompressible) 
ground state depends upon not only the total binding energy, but the 
interactions between all the charged particles which depends on $2S$.

\begin{figure}[t]
\epsfxsize=3.35in
\epsffile{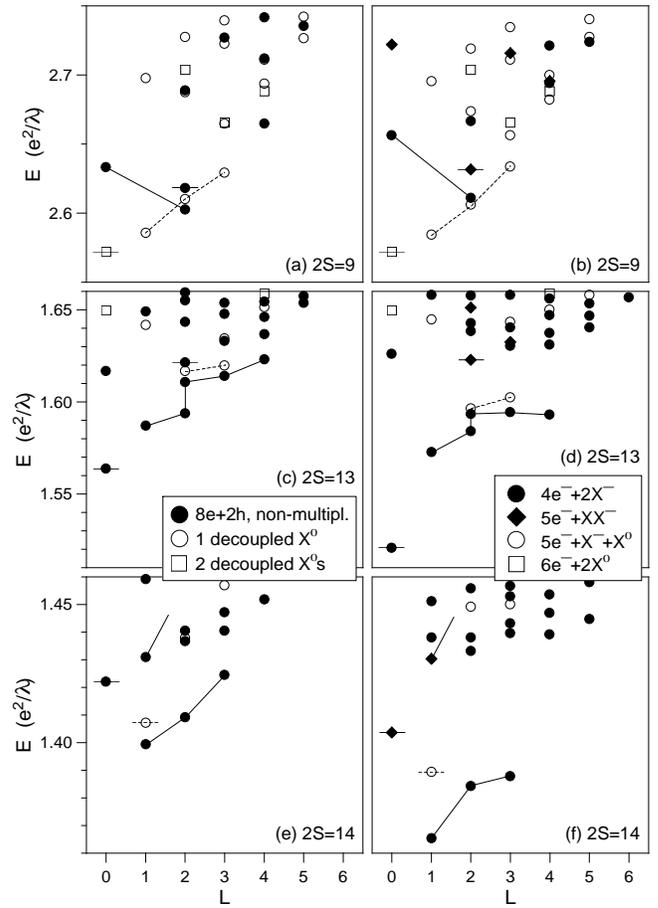}
\caption{
   Left: low energy spectra of the $8e+2h$ system on a Haldane sphere 
   at $2S=9$ (a), $2S=13$ (c), and $2S=14$ (e).
   Right: approximate spectra calculated for all possible groupings 
   containing excitons (charged composite particles interacting through 
   pseudopotentials as in Fig.~\protect\ref{fig1}).
   Lines connect corresponding states in left and right frames.
}
\label{fig2}
\end{figure}

In Fig.~\ref{fig2}, we show the low energy spectra of the $8e+2h$ system 
at $2S=9$ (a), $2S=13$ (c), and $2S=14$ (e).
Filled circles mark the non-multiplicative states, and the open circles 
and squares mark the multiplicative states with one and two decoupled
excitons, respectively.
In frames (b), (d) and (f) we plot the low energy spectra of different
charge complexes interacting through appropriate pseudopotentials (see 
Fig.~\ref{fig1}), corresponding to four possible groupings (i)--(iv).
By comparing left and right frames, we can identify low lying states of 
type (i)--(iv) in the electron-hole spectra.

In general, energies calculated from pseudopotentials $V_{AB}$ in 
Fig.~\ref{fig2} underestimate energies of the corresponding 
electron-hole system if $N$ and $2S$ are large.
This can be partially understood in terms of polarization effects in 
the two-particle pseudopotentials.
For a particular grouping and value of $2S$, it is possible to calculate 
pseudopotentials that give a very good fit to the electron-hole spectrum.
The ``correct'' pseudopotentials for the $8e+2h$ system are close to 
those of a pair of point charges with appropriate angular momenta $l_A$ 
and $l_B$, except for the hard cores.\cite{wojs2}

It is unlikely that a system containing a large number of different
species (e.g. $e^-$, $X^-$, $X_2^-$, etc.) will form the absolute
ground state of the electron-hole system.
However, different charge configurations can form low lying excited 
bands.
An interesting example is the $12e+6h$ system at $2S=17$.
The $6X^-$ grouping (v) has the maximum total binding energy 
$\varepsilon_{\rm v}=6\varepsilon_0+6\varepsilon_1$.
Other expected low lying bands correspond to the following groupings:
(vi) $e^-+5X^-+X^0$ with $\varepsilon_{\rm vi}=6\varepsilon_0+5
\varepsilon_1$ and (vii) $e^-+4X^-+X_2^-$ with $\varepsilon_{\rm vii}
=6\varepsilon_0+5\varepsilon_1+\varepsilon_2$.

\begin{figure}[t]
\epsfxsize=3.35in
\epsffile{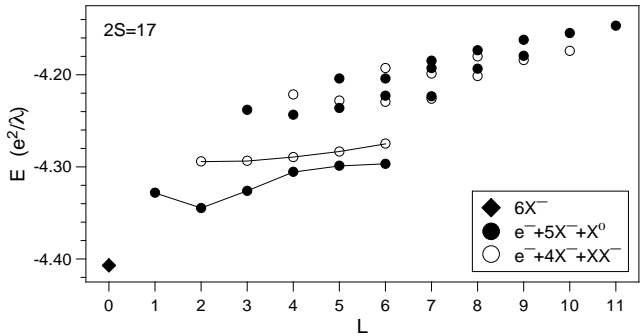}
\caption{
   Low energy spectra of different charge configurations of 
   the $12e+6h$ system on a Haldane sphere at $2S=17$: 
   $6X^-$ (diamonds), $e^-+5X^-+X^0$ (filled circles), 
   and $e^-+4X^-+X_2^-$ (open circles).
}
\label{fig3}
\end{figure}

Although we are unable to perform an exact diagonalization for the
$12e+6h$ system in terms individual electrons and holes, we can use
appropriate pseudopotentials and binding energies of groupings (v)--(vii)
to obtain the low lying states in the spectrum.
The results are presented in Fig.~\ref{fig3}.
There is only one $6X^-$ state (the $L=0$ Laughlin $\nu_{X^-}=1/3$ 
state\cite{wojs2}) and two bands of states in each of groupings 
(vi) and (vii).
A gap of 0.0626~$e^2/\lambda$ separates the $L=0$ ground state from 
the lowest excited state.

\paragraph*{Generalized Laughlin Wavefunction.}
It is known that if the pseudopotential $V({\cal R})$ decreases quickly 
with increasing ${\cal R}$, the low lying multiplets avoid (strongly 
repulsive) pair states with one or more of the smallest values of 
${\cal R}$.\cite{wojs3,wojs4}
For the (one-component) electron gas on a plane, avoiding pair states 
with ${\cal R}<m$ is achieved with the factor $\prod_{i<j}(x_i-x_j)^m$
in the Laughlin $\nu=1/m$ wavefunction.
For a system containing a number of distinguishable types of Fermions 
interacting through Coulomb-like pseudopotentials, the appropriate
generalization of the Laughlin wavefunction will contain a factor 
$\prod(x^{(a)}_i-x^{(b)}_j)^{m_{ab}}$, where $x^{(a)}_i$ is the 
complex coordinate for the position of $i$th particle of type $a$, 
and the product is taken over all pairs.
For each type of particle one power of $(x^{(a)}_i-x^{(a)}_j)$ results
from the antisymmetrization required for indistinguishable Fermions and
the other factors describe Jastrow type correlations between the 
interacting particles.
Such a wavefunction guarantees that ${\cal R}_{ab}\ge m_{ab}$, for all 
pairings of various types of particles, thereby avoiding large pair 
repulsion.\cite{halperin,haldane2}
Fermi statistics of particles of each type requires that all $m_{aa}$ 
are odd, and the hard cores defined by Eq.~(\ref{eq1}) require that 
$m_{ab}\ge{\cal R}_{ab}^{\rm min}$ for all pairs.

\paragraph*{Generalized Composite Fermion Picture.}
In order to understand the numerical results obtained in the spherical 
geometry (Figs.~\ref{fig2} and \ref{fig3}), it is useful to introduce 
a generalized CF picture by attaching to each particle fictitious flux 
tubes carrying an integral number of flux quanta $\phi_0$.
In the multi-component system, each $a$-particle carries flux $(m_{aa}-1)
\phi_0$ that couples only to charges on all other $a$-particles and 
fluxes $m_{ab}\phi_0$ that couple only to charges on all $b$-particles,
where $a$ and $b$ are any of the types of Fermions.
The effective monopole strength \cite{chen,jain,wojs3,sitko} seen by 
a CF of type $a$ (CF-$a$) is
\begin{equation}
   2S_a^*=2S-\sum_b(m_{ab}-\delta_{ab})(N_b-\delta_{ab})
\label{eq2}
\end{equation}
For different multi-component systems we expect generalized Laughlin 
incompressible states (for two components denoted as $[m_{AA},m_{BB},
m_{AB}]$) when all the hard core pseudopotentials are avoided and CF's 
of each kind fill completely an integral number of their CF shells 
(e.g. $N_a=2l_a^*+1$ for the lowest shell).
In other cases, the low lying multiplets are expected to contain different 
kinds of quasiparticles (QP-$A$, QP-$B$, \dots) or quasiholes (QH-$A$, 
QH-$B$, \dots) in the neighboring incompressible state.

Our multi-component CF picture can be applied to the system 
of excitonic ions, where the CF angular momenta are given by 
$l_{X_k^-}^*=|S_{X_k^-}^*|-k$.
As an example, let us first analyze the low lying $8e+2h$ states 
in Fig.~\ref{fig2}.
At $2S=9$, for $m_{e^-e^-}=m_{X^-X^-}=3$ and $m_{e^-X^-}=1$ we predict 
the following low lying multiplets in each grouping:
(i) $2S_{e^-}^*=1$ and $2S_{X^-}^*=3$ gives $l_{e^-}^*=l_{X^-}^*=1/2$.
Two CF-$X^-$'s fill their lowest shell ($L_{X^-}=0$) and we have two 
QP-$e^-$'s in their first excited shell, each with angular momentum 
$l_{e^-}^*+1=3/2$ ($L_{e^-}=0$ and 2).
Addition of $L_{e^-}$ and $L_{X^-}$ gives total angular momenta $L=0$ 
and 2.
We interpret these states as those of two QP-$e$'s in the incompressible 
[331] state.
Similarly, for other groupings we obtain:
(ii) $L=2$;
(iii) $L=1$, 2, and 3; and
(iv) $L=0$ ($\nu=2/3$ state of six electrons).

At $2S=13$ and 14 we set $m_{e^-e^-}=m_{X^-X^-}=3$ and $m_{e^-X^-}=2$ 
and obtain the following predictions.
First, at $2S=13$:
(i) The ground state is the incompressible [332] state at $L=0$;
the first excited band should therefore contain states with one QP-QH 
pair of either kind.
For the $e^-$ excitations, the QP-$e^-$ and QH-$e^-$ angular momenta 
are $l_{e^-}^*=3/2$ and $l_{e^-}^*+1=5/2$, respectively, and the allowed
pair states have $L_{e^-}=1$, 2, 3, and 4.
However, the $L=1$ state has to be discarded, as it is known to have high 
energy in the one-component (four electron) spectrum.\cite{sitko}
For the $X^-$ excitations, we have $l_{X^-}^*=1/2$ and pair states can
have $L_{X^-}=1$ or 2.
The first excited band is therefore expected to contain multiplets at
$L=1$, $2^2$, 3, and 4.
The low lying multiplets for other groupings are expected at:
(ii) $L=2$ and 3;
(iii) $2S_{X_2^-}^*=3$ gives no bound $X_2^-$ state; setting 
$m_{e^-X^-}=1$ we obtain $L=2$; and 
(iv) $L=0$, 2, and 4.
Finally, at $2S=14$ we obtain:
(i) $L=1$, 2, and 3;
(ii) incompressible [3*2] state at $L=0$ ($m_{X^-X^-}$ is irrelevant 
for one $X^-$) and the first excited band at $L=1$, 2, 3, 4, and 5;
(iii) $L=1$; and
(iv) $L=3$.

For the $12e+6h$ spectrum in Fig.~\ref{fig3} the following CF predictions
are obtained:
(v) For $m_{X^-X^-}=3$ we obtain the Laughlin $\nu=1/3$ state with $L=0$.
Because of the hard core of $V_{X^-X^-}$, this is the only state of
this grouping.
(vi) We set $m_{X^-X^-}=3$ and $m_{e^-X^-}=1$, 2, and 3.
For $m_{e^-X^-}=1$ we obtain $L=1$, 2, $3^2$, $4^2$, $5^3$, $6^3$, $7^3$, 
$8^2$, $9^2$, 10, and 11.
For $m_{e^-X^-}=2$ we obtain $L=1$, 2, 3, 4, 5, and 6.
For $m_{e^-X^-}=3$ we obtain $L=1$.
(vii) We set $m_{X^-X^-}=3$, $m_{e^-X_2^-}=1$, $m_{X^-X_2^-}=3$, and 
$m_{e^-X^-}=1$, 2, or 3.
For $m_{e^-X^-}=1$ we obtain $L=2$, 3, $4^2$, $5^2$, $6^3$, $7^2$, $8^2$, 
9, and 10.
For $m_{e^-X^-}=2$ we obtain $L=2$, 3, 4, 5, and 6.
For $m_{e^-X^-}=3$ we obtain $L=2$.
In groupings (vi) and (vii), the sets of multiplets obtained for higher 
values of $m_{e^-X^-}$ are subsets of the sets obtained for lower values, 
and we would expect them to form lower energy bands since they avoid 
additional small values of ${\cal R}_{e^-X^-}$.
However, note that the (vi) and (vii) states predicted for $m_{e^-X^-}=3$ 
(at $L=1$ and 2, respectively) do not form separate bands in 
Fig.~\ref{fig3}.
This is because the $V_{e^-X^-}$ pseudopotential increases more 
slowly than linearly as a function of $L(L+1)$ in the vicinity of 
${\cal R}_{e^-X^-}=3$ (see Fig.~\ref{fig1}).
In such case the CF picture fails.\cite{wojs3}

The agreement of our CF predictions with the data in Figs.~\ref{fig2}
and \ref{fig3} (marked with lines) is really quite remarkable and 
strongly indicates that our multi-component CF picture is correct.
We were indeed able to confirm predicted Jastrow type correlations in 
the low lying states by calculating their coefficients of fractional 
parentage.\cite{wojs3,shalit}
We have also verified the CF predictions for other systems that we were 
able to treat numerically.
If exponents $m_{ab}$ are chosen correctly, the CF picture works 
well in all cases.

\paragraph*{Summary.}
Charged excitons and excitonic complexes play an important role in 
determining the low energy spectra of electron-hole systems in a strong 
magnetic field.
We have introduced general Laughlin type correlations into the 
wavefunctions, and proposed a generalized CF picture to elucidate 
the angular momentum multiplets forming the lowest energy bands for
different charge configurations occurring in the electron-hole system.
We have found Laughlin incompressible fluid states of multi-component 
plasmas at particular values of the magnetic field, and the lowest 
bands of multiplets for various charge configurations at any value 
of the magnetic field.
It is noteworthy that the fictitious Chern--Simons fluxes and charges 
of different types or colors are needed in the generalized CF model.
This strongly suggests that the effective magnetic field seen by the 
CF's does not physically exist and that the CF picture should be 
regarded as a mathematical convenience rather than physical reality.
Our model also suggests an explanation of some perplexing observations 
found in photoluminescence, but this topic will be addressed in a separate 
publication.

We thank P. Hawrylak and M. Potemski for helpful discussions.
AW and JJQ acknowledge partial support from the Materials Research 
Program of Basic Energy Sciences, US Department of Energy.
KSY acknowledges support from the Korea Research Foundation (Project
No. 1998-001-D00305).

\end{document}